\documentstyle[twoside,fleqn,espcrc2,psfig]{article}

\title{D-brane decay and Hawking Radiation\footnote{Talk given at
Strings 97}}

\author{Sumit R. Das%
\address{ Tata Institute of Fundamental Research \\
          Homi Bhabha Road \\
          Mumbai 400 005. India.}}

\begin{document}

\begin{abstract}
Tree level decay amplitudes of near-BPS D-brane configurations are
known to exactly reproduce Hawking radiation rates from corresponding
black holes at low energies even though the brane configurations
describe semiclassical black holes only when the open string couplings
are large.  We show that a large class of one (open
string) loop corrections to emission processes from D-branes vanish at
low energies and nonvanishing loop contributions have an energy
dependence consistent with black hole answers, thus providing a
justification for the agreement of the tree level results with
semiclassical answers.
\end{abstract}

\maketitle


\section{Introduction}

Recently the idea that massive string states become black holes when
the coupling is large \cite{sussk,duff,sen} has been very successful.
In particular, five dimensional extremal black holes with large
horizons are described by bound states of D-branes whose degeneracies
exactly reproduce the Beckenstein-Hawking entropy \cite{stromvafa}
This result was extended to four dimensional
extremal black holes \cite{fdh} and to
spinning black holes \cite{spin}.

The entropy and hawking temperature continue to agree in the near
extremal limit
\cite{callanmal,horostrom}.
It was also found that the lowest order decay rate of slightly
nonextremal D-brane configurations is proportional to the horizon area
\cite{callanmal} which is consistent with the semiclassical Hawking
radiation from such holes, as shown in
\cite{dmw}.
Rather surprisingly the two emission (and
absorption) rates for neutral scalars 
were in fact found to agree {\it exactly} 
\cite{dmatb}. Exact
agreements were also found for neutral and charged scalars in five and
four dimensional black holes \cite{gubkle}.  Even more remarkably, the
grey body factors which describe a nontrivial energy dependence of the
absorption cross-section at higher energies are also in exact
agreement
\cite{malstrom}, a result verified in the four dimensional case as
well \cite{gubkleb}.A general analysis of classical
absorption by such black holes and the possibility of agreement of 
classical and D-brane greybody factors, has been carried out in
\cite{dealwis}.

A more detailed test of these ideas is provided by the emission and
absorption of certain ``fixed'' scalars by five dimensional black
holes \cite{fixed}, where agreement between the D-brane and classical
calculations were demonstrated in \cite{callanfixed}. The grey body
factors at higher energies agree as well \cite{klebkras}.

A related example where there seem to be exact agreement of D-brane
and general relativity results is the absorption of $l =0$ and $l = 1$
waves by extremal 3-branes with no momentum \cite{igorc,igore}.  The
cross-sections for higher partial waves agree upto numerical factors
\cite{igore}. In this case the near-extremal entropy also differs by a
numerical factor \cite{igord}.

For systems arbitrarily far from extremality (like the Schwarzschild
black hole) it has been argued that the microscopic and semiclassical
answers should match only at a special value of the coupling at which
the horizon curvature is of the string scale \cite{sussk,horpol}.
It was shown in \cite{horpol} that in all known cases the stringy and
semiclassical entropies indeed match at this point, upto numerical
factors.

\section{The puzzle}

In a sense these spectacular results are puzzling. In the standard
semiclassical description of Hawking radiation the decay rate into
particles of energy-momentum $(\omega,k)$ is given by
\begin{equation}
\Gamma(\omega) = {\sigma (\omega) \over e^{\beta_H \omega}\pm 1}
{d^dk \over (2\pi)^d}
\label{eq:neone}
\end{equation}
where $d$ denotes the number of spatial dimensions and $\sigma(\omega)$
denotes the absorption cross-section of waves of frequency $\omega$
by the black hole. $\sigma (\omega)$ thus encodes the space-time
structure of the black hole.

On the other hand in D-brane perturbation theory Hawking radiation
appears as a result of transitions between solitonic states in {\em
flat} space-time, usually in the lowest order in string perturbation
theory. Thermality of the radiation is due to a large degeneracy of
initial states of the brane system.

These are quite different pictures. Yet at low energies the answers for
$\Gamma (\omega)$ agree in the cases mentioned above.

In fact the semiclassical black hole picture and the perturbative
D-brane picture are descriptions of the same object in two quite
different regimes. D-brane states are expected to describe 
semiclassical black holes
when $(gQ)$ is large, where $g$ is the string coupling and $Q$ is a
typical charge of the hole. This product $(gQ)$ is in fact the open
string coupling. The D-brane calculations are, however, performed at
weak open string coupling. For extremal BPS states there are
well-known nonrenormalization theorems which ensure that the
degeneracy of states do not change as we increase the coupling. But
for non-BPS states there are no such obvious theorems.

To appreciate the point consider the metric for the five dimensional near
extremal black hole which is described by bound states of 1D-branes and
5D-branes with some momentum flowing along the 1D brane. Non extremality
is introduced by allowing both left and right momenta on the 1D brane
and the classical solution of the low energy effective action has 
a ten-dimensional string metric
\begin{eqnarray}
ds^2 & = & (1+{r_1^2 \over r^2})^{-1/2}(1+{r_5^2 \over r^2})^{-1/2}[
dt^2 - dx_5^2 \nonumber \\
& & -{r_0^2 \over r^2}({\rm cosh}\sigma~dt + {\rm
sinh}\sigma
~dx_5)^2]\nonumber \\
& &- (1+{r_1^2 \over r^2})^{1/2}(1+{r_5^2 \over r^2})^{-1/2}
[dx_1^2 + dx_2^2 \nonumber \\
& & + dx_3^2 + dx_4^2] \nonumber \\
& &- (1+{r_1^2 \over r^2})^{1/2}(1+{r_5^2 \over r^2})^{1/2}
[(1 - {r_0^2 \over r^2})^{-1} dr^2 \nonumber \\
& &+ r^2 d\Omega_3^2]
\label{eq:bone}
\end{eqnarray}
The various length scales are given in terms of the charges by
\begin{eqnarray}
 r_1^2 & = & {16 \pi^4 \alpha '^3 (g Q_1) \over V} \nonumber \\ r_5^2
 & = & \alpha ' (gQ_5)\nonumber \\ {1 \over 2} r_0^2 {\rm sinh}
 2\sigma & = & { 16 \pi^4 \alpha '^4 (g^2 N) \over R^2 V} \nonumber \\
 r_N^2 & = & r_0^2 {\rm sinh}^2 \sigma
\label{eq:btwo}
\end{eqnarray}
Here $\alpha ' = 1/(2\pi T)$ where $T$ is the elementary string
tension, $g$ is the string coupling. The brane configuration lies on a
$T^4 \times S^1$ with the one brane along the $S^1$. The radius of
this $S^1$ is $R$, while the volume of the $T^4$ is $V$. The integers
$Q_1, Q_5, N$ are the 1-brane RR charge, 5-brane RR charge and the
total momentum.  The extremal limit is $r_0 \rightarrow 0$ and $\sigma
\rightarrow \infty$ with $N$ held fixed.

It is clear from the classical solution that
the classical limit of the string theory corresponds to $g \rightarrow 0$
with $ gQ_1, gQ_5, g^2 N$ held fixed \cite{malstrom}.  In fact we have
large black holes (compared to string scale) when
$  gQ_1, gQ_5, g^2 N > 1 $ and small holes when 
$  gQ_1, gQ_5, g^2 N < 1$. It is in the latter regime that the D-brane
description in terms of a bound state of 1D and 5D branes
with some momentum along the 1D brane is reliable.

In the dilute gas regime $r_N << r_1, r_5$ the size of the black hole
is controlled by $(gQ_1)$ and $(gQ_5)$ which are the effective open
string coupling constants.
The full classical
solution can be obtained by summing over an infinite number of string
diagrams which does not contain any closed string loop, but contains
all terms with closed strings terminating on an aribtrary number of
branes. Each such insertion carries a factor $gQ$ which has to be held
finite. In other words we have to sum over all open string loops.
Closed string loops do not contain any factor of the charge $Q$ and
are therefore suppressed.  This perturbation expansion is a
description of the black hole expanded around flat space-time with the
curvature emerging as a result of summing over open string loops.

Another example which we will consider in the following is the
self-dual 3-brane in Type IIB theory, first considered in this context
in \cite{igord,igorc,igore}. The extremal solution has a zero horizon
area, but is completely {\it nonsingular} and the dilaton is a
constant in the classical solution. The extremal string metric is
given by
\begin{eqnarray}
ds^2 & = & A^{-1/2}(-dt^2 + dx_1^2 + dx_2^2 + dx_3^2) \nonumber \\
& & + A^{1/2}(dr^2 + r^2 d\Omega_5^2)
\label{eq:btwob}
\end{eqnarray}
with
\begin{equation}
A(r) = 1 + {4\pi \alpha '^2 (gN) \over r^4}
\label{eq:btwoc}
\end{equation}
where $N$ is the RR charge. The curvature at the horizon $r = 0$ is
$\sim 1/[\alpha '{\sqrt{gN}}]$. Thus when $gN$ is large, the
curvatures are small and one may trust the supergravity limit. For
$gN << 1$ this system is well described by parallel 3D branes. A
great advantage of using extremal 3-branes is that the weak coupling
description is well known in terms of Dirichlet open string theory,
while for the five or four dimensional black holes not much is
known about the properties of the bound states.

The question is : why is it that the absorption or emission cross
sections calculated in tree level open string theory agree in detail
with the semiclassical black hole answers. Why is it that open string
loops do not alter the result.

In this talk I will summarize some results which seem to offer
an answer to this question. Some of these results are described
in \cite{dasloop}. The results of Section 7 are however new and
not published anywhere else. For further results concerning loop
corrections see the contribution of Klebanov to these proceedings
\cite{klebast}.

\section{The issue of loop corrections}

A little thought shows that the situation is not as puzzling as it
first appears. Consider for example absorption by extremal 
black holes which have a single
length scale in the problem.  Examples are fat black holes with
$r_1 = r_5 = r_N = R$ or extremal parallel branes, like 3-branes.
Let us call this length scale $l$.
In the classical solution the
string coupling can enter only through this length scale $l$ which is
typically given by the form
\begin{equation}
l^{(d-3)} \sim g Q \alpha '^{(d-3/2)}
\label{eq:addone}
\end{equation}
where $d$ denotes the number of non-compact dimensions. It is then
clear that the classical absorption cross-section has to be of the
form
\begin{equation}
\sigma_{class} \sim l^{d-2} F(\omega^{(d-3)}g Q \alpha '^{(d-3/2)})
\label{eq:addtwo}
\end{equation}
On general grounds we expect that this classical answer should agree
with the D-brane answer when $gQ$ is large. However the above
expression shows that for sufficiently small $\omega$ one may have the
factor $\omega^{(d-3)}g Q \alpha '^{(d-3/2)}$ small even if $gQ$ is large
so that one may imagine performing a Taylor expansion of the function
$F$, which then becomes a power series expansion in the string
coupling $g$ as well \cite{igorc}. 
The spectacular success of the tree level D-brane calculations of the
absorption cross-section then means that the {\it lowest order} term
in this expansion has been shown to agree with the {\it lowest order}
term in D-brane open string perturbation theory.
 
The puzzle regarding this agreement of D-brane and classical
calculations may be now restated as follows : In the classical limit a
higher power of the string coupling comes with a higher power of the
energy in a specific way dictated by (\ref{eq:addtwo}). On the other
hand, on the D-brane side these higher powers of coupling are to be
obtained in open string perturbation theory and there is no {\it a
  priori} reason why this should also involve higher powers of energy
in precisely the same way.

This implies that the following two kinds of loop corrections must be
absent for the correspondence to work. (1) For a given absorption
or emission process open string loop
diagrams with the same external states must be suppressed at low
energies compared to the tree diagram. (2) Suppose we concentrate on
emission of some given closed string state and let the leading order
tree process give a cross-section $\sigma \sim g^\alpha$ with some
energy dependence. For $\alpha$ large enough it is possible that there
is a string loop process with the same dependence on the coupling
\footnote{The importance of this was emphasized to me by I. Klebanov.}. 
This must, of course, involve external states which are {\it
different} from the tree process. For processes where the D-brane
tree level calculation gave the correct answer such loop corrections
must be suppressed at low energies.

In fact the black hole correspondence demands more. Nonzero higher
loop effects must appear in the precise combination of the string
coupling and the energy displayed in (\ref{eq:addtwo}).

Note that in the D-brane calculations higher powers of $\omega$ may
appear with the same power of $g$ through grey body factors arising
from the thermal distributions accompanying the initial states.
Nevertheless the black hole answer tells us how the D-brane
perturbation series should look like.

Additional evidence for this picture appears from an application
of the correspondence principle of \cite{horpol} to near-extremal
black holes made of parallel $p-D$-branes with no other charge. 
For such black holes the entropy due to the gas of massless modes
on the p-brane worldvolume agrees, upto a numerical factor, with the
black hole entropy at the correspondence point defined as the value
of the string coupling where the string metric curvature at the
horizon becomes of the order of the string scale. Now, it is known
that for any spherically symmetric static black hole in any number
of dimensions the absorption cross-section of massless minimally
coupled scalars is exactly equal to the horizon area in the low
energy limit \cite{dgm}. However, it turns out that with fairly mild
assumptions about the nature of worldvoulme interactions the
pD-brane cross-sections fail to reproduce this lowest order answer
except for 1-branes and 3-branes \cite{daspbrane}. This result may
be interpreted to imply that the agreement of absorption/emission
cross-sections should be understood only in the sense of a
coupling-energy expansion.

Maldacena \cite{Maldacenab} has derived non-renormalization properties
of the low energy Yang-Mills field theory on the brane to explain
agreement of the non-extremal entropy with the black hole answers.
Similar arguments apply to some of the absorption/emission processes.
We will, however, need properties of some higher order terms in the
Born-Infeld action of branes, like terms involving products of four
open string field derivatives. Such terms have a specific stringy 
origin and cannot be investigated in the Yang-Mills approximation.
Furthermore, as mentioned above, we not only need to show that 
certain higher loop terms vanish but also need to show that the
nonvanishing higher loop contributions are consistent with the
black hole answer.

Consequently we look at properties of open string loop diagrams
explicitly. Strictly speaking, this restricts us to a study of
parallel branes - since in that case we know the microscopic theory
accurately in terms of Chan-Paton factors. Our results would be
rigourously valid for the extremal 3-brane. For the five and four
dimensional black holes enough is not known about the bound states
involved to enable a reliable loop calculation. However we do expect
some consequences of our results for these black holes as well - this
is based on the success of the effective long string picture.

\section{Two open string processses}

Consider the process of emission of a massless closed string state
from two open strings annahilating on the worldvolume of $N$ parallel
p-D branes. One example is the emission of scalars in the five
dimensional black hole which arise from components of the ten
dimensional graviton $h_{IJ}$ along the $T^4$ direction orthogonal to
the 1D brane. The relevant interaction term in the tree level
effective lagrangian density is given by 
\begin{equation} {\sqrt 2} \kappa
h_{IJ}(x^I,x^\alpha,x^i)\partial_\alpha X^I \partial^\alpha X^J
\label{eq:netwo}
\end{equation}
Here $I,J$ denote directions on the $T^4$ orthogonal to the 1D brane,
$\alpha,\beta$ denote either time or the 1D brane direction and
$i,j$ denote directions transverse to $T^4 \times S^1$ on which the
branes are wrapped. Another example is the absorption of scalars
arising from the longitudinal components of the ten dimensional
gravition by a collection of 3-branes. The interaction term in the
low energy effective
action is then given by
\begin{equation}
h_{\alpha \beta}{\rm Tr}[F_\alpha^\gamma F_{\gamma\beta}
-{1\over 2}F^2 
+ \partial_\gamma X^i \partial^\gamma X^i +\cdots]
\label{eq:nethree}
\end{equation}
where $\alpha,\beta,\gamma$ denote worldvolume indices and $i,j$
transverse indices and $F$ is the gauge field strength. The elipses
denote the fermionic terms.

For emission into $S$-waves the closed string fields $h_{IJ}$ in
(\ref{eq:netwo}) and $h_{\alpha\beta}$ in (\ref{eq:nethree}) are
independent of the transverse coordinates. The lowest energy 
contribution comes from $h_{IJ}, h_{\alpha\beta}$ which are dependent
only on the worldvolume coordinates. These terms can be in fact read
off from the effective action {\em in the absence of closed string 
fields}. This follows from the principle of equivalence. For example
the flat space term for the 1D-5D system is
\begin{equation}
\delta_{IJ}\partial X^I \partial X^J
\label{eq:nefour}
\end{equation}
In the presence of a nontrivial metric $\delta_{IJ}$ has to be
replaced by a tensor in the space transverse to the 1D brane
but longitudinal to the 5D brane. Thus one may have
\begin{equation}
\delta_{IJ} \rightarrow (G_{IJ} + R_{IJ} + \cdots)
\label{eq:nefive}
\end{equation}
where $G_{IJ}$ is the metric and $R_{IJ}$ is the Ricci etc.
In (\ref{eq:nesix}) the only term which does not involve derivatives
of $h_{IJ}$ is the metric $G_{IJ}$ itself. All the other terms
are therefore suppressed at low energies. This is however the term
which gives rise to (\ref{eq:netwo}).
Terms responsible for emission of higher partial waves cannot be
read off from the flat space term in this fashion, since in this
case one has to consider dependence on the transverse coordinates
\cite{igorc,igord}.

We want to compute the one (open string) loop correction to these
terms. Thus we have to compute an annulus diagram with two open string
massless vertices on the boundary and a massless closed string vertex
in the interior of the annulus. The coordinate fields transverse to
the brane satisfy Dirichlet conditions on the boundaries while the
longitudinal ones satisfy Neumann conditions.  As we have just argued
for the lowest energy contribution it is sufficient to evaluate these
terms {\em without} the closed string insertion.  Since we have an
{\em oriented} open string theory on the brane we have both a planar
contribution ${\cal A}_p$ as well as a non planar contribution ${\cal
A}_{np}$ which are given by 
\begin{eqnarray} 
{\cal A}_p & = & K_{(2,0)} {\rm
Tr}[V(k) \Delta V(-k) \Delta] \nonumber \\ {\cal A}_{np} & = &
K_{(1,1)} {\rm Tr}[V(k) \Omega \Delta V(-k) \Omega \Delta]
\label{eq:nesix}
\end{eqnarray}
where $K_{(2,0)}$ and $K_{(1,1)}$ are Chan Paton factors, $\Delta$
is the open string propagator and $\Omega$ is the twist operator.
The net contribution is
\begin{equation}
{\cal A} = {\cal A}_p + {\cal A}_{np}
\label{eq:neseven}
\end{equation}
However both the terms in (\ref{eq:neseven}) vanish individually
due to supersymmetry of the background. This may be seen for example
in the Green-Schwarz formalism in the light cone gauge. To apply this
formalism to D-branes one has to perform a double Wick rotation as
explained in \cite{GREEN}.The boundary conditions for the
Green-Schwarz fermions $S^a$ are obtained by requiring that half of the
supersymmetries are preserved and become 
\begin{equation} 
S^a = M^a_b S^b
\label{eq:hone}
\end{equation}
where $M$ is a matrix satisfying the conditions \cite{grav,garmyers}
\begin{eqnarray}
M^T M & = & I \nonumber \\
M^T \gamma^I M & = & - \gamma^I~~~~~~I : {\rm Dirichlet} \nonumber \\
M^T \gamma^\alpha M & = & \gamma^\alpha~~~~~~~\alpha : {\rm Neumann}
\label{eq:htwo}
\end{eqnarray} 
The vertex operators for the open string massless states with
polarizations in the Dirichlet and Neumann directions are given by
(respectively)
\begin{eqnarray}
V_D & = & \zeta_I (k) (\partial_\sigma X^I - S_+ \gamma^{I\alpha} S_+
 k_\alpha)e^{ikX}\nonumber \\ V_N & = & \zeta_\beta (k)(\partial_\tau
 X^\beta - S_+ \gamma^{\beta\alpha} S_+ k_\alpha)e^{ikX}
\label{eq:hthree}
\end{eqnarray}
The trace involves a trace of the zero modes of the fermions $S^a_0$
and we require at least eight fermionic fields to yield a nonzero
trace. However the annulus diagram with only two open string insertions
contains at most four fermionic fields and therefore they vanish. 
This means that there is no $O(g^2\omega^2)$ contribution to the
amplitude.

The first nonzero contribution comes when the closed string insertion
is taken into account since each closed string vertex contains four
fermion fields (two from the left sector and two from the right
sector). So with two open strings and one closed string there are
the required eight fermionic fields. Clearly, to evaluate this term
we may ignore the terms in the vertex operators which contain the
bosonic fields. Furthermore in the lowest energy contribution we may
ignore the $e^{ikX}$ parts of the vertex operators since they will
result in factors of the momentum. Thus any nonzero contribution
to the amplitude from such a term would be of order $O(g^2 \omega^4)$.
We will return to the implication of such a nonzero term later.

The above result may be directly applied to the absorption by extremal
3-branes with zero momentum. It is clear from (\ref{eq:nethree}) that
the tree level amplitude is of order $O(g \omega^2)$. Converting this
to a transition rate and summing over initial states leads to a
cross-section for $S$-waves
\begin{equation}
\sigma_{3D}^{l=0, tree} \sim (gN)^2 \omega^3
\label{eq:neeight}
\end{equation}
and as shown in \cite{igorc} the coefficient is in exact agreement
with the semiclassical cross-section. If there was a nonzero
contribution to the amplitude of order $O(g^2\omega^2)$ one would
have a $(gN)^4 \omega^3$ term in the cross-section in contradiction
with the classical cross-section. Thus our result ``explains'' why
the perturbative D-brane calculation yields the low energy classical
result.

\section{Three open string processes}

Processes involving three open strings are relevant for the absorption of
$l = 1$ modes by the extremal 3-brane through terms like
\begin{equation}
(\partial_i \phi){\rm Tr}[X^i F^2]
\end{equation}
Using arguments similar to that in the previous section it may be easily
seen that the lowest order one loop term could be of the order
$O(g^{5/2} \omega^4)$. The resulting cross-section may be seen to be
suppressed in energy compared to the tree level contribution.

\section{Four open string processes}

Processes involving four open strings display for the first time
nontrivial effects of the sum over planar and non-planar
diagrams. These processes are relevant for emission of ``fixed
scalars'' in the five dimensional black hole
\cite{callanfixed,klebkras}.  In the effective long string model,
where the 1D brane has to be considered multiply wound
\cite{maldasuss} along the lines of \cite{dmata,mathurd},
expansion of the Born-Infeld action in the static gauge with flat
worldsheet yields an action 
\begin{eqnarray} 
S & \sim & \int dx^0 dx^5
e^{-\phi}[ 1 + {1\over 2}e^{\phi/2} G_{IJ}^E \partial_+ X^I \partial_-
X^J \nonumber \\
& &  -{1\over 8} e^\phi
G^E_{IJ}G^E_{KL}\partial_+X^I\partial_+X^J\partial_-X^K \partial_-X^L
\nonumber \\
& & +\cdots]
\label{eq:nenine}
\end{eqnarray} 
where $G^E$ denotes the Einstein frame metric. An example of a
fixed scalar is the size of the $T^4$, $G_{IJ} =
e^{2\nu}\delta_{IJ}$. If we require that the five dimensional
dilaton $\phi_5 = \phi - 2\nu$ is not emitted, then it is seen that
the lowest order term in (\ref{eq:nenine}) is quartic in $X^I$. When
no $h_{55}$ or $h_{5I}$ are emitted and $Q_1 = Q_5$ the tree level
cross-section agrees with the classical answer \cite{callanfixed}. For
$Q \neq Q_5$ one cannot consistently set $h_{55}$ to zero and there is
no agreement \cite{klebkrasb}, while in the presence of $h_{5I}$
agreement can be obtained by modifying the effeactive action
\cite{klrats}. We will restrict our attention to the case of $h_{55} =
0$ and $Q_1 = Q_5$. In this case the lowest energy interaction may be
read off from the effective action in the absence of any closed string
field along the lines of the previous section.

The one loop contribution is from an annulus diagram with four open
string vertices at the boundaries. Now we have the required number of
fermion zero modes to give a nonzero answer since each open string
vertex operator contains two fermions and there are eight fermions in
all.  Thus the lowest order nonzero contribution comes from the term
in which we can ignore the fermions in the open string propagators,
ignore the bosonic parts of the vertex operators and ignore the
$e^{ikX}$ parts as explained above. Furthermore we may replace the
fermionic fields by their zero modes so that there are no
oscillators. If ${\cal A}_{m,n}$ denotes the amplitude with $m$
vertices on one boundary and $n$ vertices on the other boundary, the
various contributions are 
\begin{eqnarray} 
{\cal A}_{(4,0)} & = & 2 K_{(4,0)} {\rm
Tr}[V(k_1)\Delta V(k_2) \Delta V(k_3)
\nonumber \\
& & \Delta V(k_4)\Delta] \nonumber \\ 
{\cal A}_{(3,1)}  & = & 8 K_{(3,1)} {\rm Tr}[V(k_1)\Omega \Delta V(k_2)
\nonumber \\
& & \Omega\Delta V(k_3)\Delta V(k_4)\Delta] \nonumber \\ 
{\cal A}_{(2,2)}  & = & 6
K_{(2,2)} {\rm Tr}[V(k_1)\Omega \Delta V(k_2)
\nonumber \\
& & \Delta V(k_3) \Omega \Delta
V(k_4)\Delta]
\label{eq:neeleven}
\end{eqnarray} 
The effect of a twist operator is to change signs of nonzero
oscillators in the following way 
\begin{equation} 
\alpha_n \rightarrow (-1)^n
\alpha_n~~~~~S_n \rightarrow (-1)^n S_n 
\end{equation} 
However to this order
there are no oscillators ! Thus the traces above are all the same -
the only difference being in the Chan Paton factors.

A remarkable cancellation happens for single branes. Now the gauge
group is $U(1)$ and 
\begin{equation} 
K_{(4,0)} = - K_{(3,1)} = K_{(2,2)} = e^4 
\end{equation}
so that the sum of the three contributions in (\ref{eq:neeleven})
vanishes.  This cancellation is similar to the cancellation between
different topologies for the one loop contribution to $F^4$ term in
the Type I superstring \cite{Tseytlina} and is possibly related to
similar non-renormalizations required in M(atrix) theory \cite{BFSS}.
To the extent we can trust the single long effective string model,
this may be taken to be an explanation of why the fixed scalar
cross-section is correctly reproduced. The cases for nonvanishing
$h_{55}$ or $h_{5I}$ cannot be treated in this manner since these
interaction terms cannot be read off from the flat space effective
action using principle of equivalence.

\section{How big are the loop effects ?}

As we have discussed the D-brane-black hole correspondence requires
more than vanishing of certain terms in the open string loop
corrections to absorption or emission processes : the form of the
classical cross-section gives the precise form for these loop
corrections for the case where there is only a single length scale
in the problem. In this section we investigate this issue for the
case of absorption by 3-branes where the only length scale is 
given by $l \sim (gN)^{1/4} {\sqrt \alpha '}$. 

Since the tree level result is $\sim (gN)^2 \omega^3$ the expansion of
the classical cross-section following from (\ref{eq:addtwo}) then
shows that the one loop correction should be of the form
\begin{equation}
\sigma_{3D-1 loop}\sim (gN)^4 \omega^{11}
\end{equation}

In Section 4. we found that, on the basis of zero mode counting, there
is a possibility that there is a one-loop amplitude involving two open
string states of the form $g^2 \omega^4$. This would lead to a
cross-section 
\begin{eqnarray}
\sigma & \sim & {1\over \omega}\int \prod_{i=1}^2{d^3 p^{(i)} 
\over |p^{(i)}|}
\delta^4(p^{(1)}+p^{(2)}-k) (g^4 \omega^8) 
\nonumber \\
& \sim & g^4 \omega^7
\label{eq:nethirteen}
\end{eqnarray}
where $(p^{(1)},p^{(2)})$ denote the momenta of the open strings which
are in the Neumann directions and $k$ denotes the momentum of the
closed string massless state which may be in any direction. If such a
term was indeed present we would have a direct contradiction of the
correspondence of 3-branes with seven dimensional extremal black
holes. To examine this we need to calculate the coefficient of the
term.

We will take the polarizations of the open string states $\eta_1^\mu,
\eta_2^\mu$ to be in any of the directions (the longitudinal
polarizations coreespond to worldvolume gauge fields while the
transverse polarizations are the scalars). The polarization of the
massless closed string state $\epsilon_{\alpha\beta}$
is purely longitudinal and traceless - these are the minimally
coupled scalars in the seven dimensional theory whose cross-sections
have been computed in \cite{igorc,igore}. The absorbed scalar is also
neutral which means that the spatial components of $k^\mu$ are
entirely in the transverse directions.

As discussed above, for the lowest energy contribution, the vertex
operators may be simplified. The open string operators may be replaced
by 
\begin{equation} 
(S_0 \gamma^{\mu\alpha} S_0)p^{(m)}_\alpha \eta^{(m)}_\mu
\label{eq:nefourteen}
\end{equation}
where the index $m = 1,2$ labels the open string. The indices $\mu,\nu$
may run over all the ten directions, while the indices $\alpha,\beta$
runs over the four worldvolume directions. The closed string vertex is
similarly replaced by
\begin{equation}
(S_0 \gamma^{\alpha\mu}S_0)(S_0 \gamma^{\beta\nu}S_0)
\epsilon_{\alpha\beta}k_\mu k_\nu
\label{eq:nefifteen}
\end{equation} 
The final answer for the planar diagram is proportional to the
well known trace \cite{GSW} 
\begin{equation} 
t^{ijklmnpq} = {\rm
Tr}[R_0^{ij}R_0^{kl}R_0^{mn}R_0^{pq}] 
\end{equation} 
where 
\begin{equation} 
R_0^{ij} = S_0\gamma^{ij} S_0 
\end{equation} 
The trace is evaluated e.g. in
\cite{GSW}. Using the restrictions on the polarizations and momenta
components described above it may be seen that there could be terms
like 
\begin{equation} 
(p^{(1)}_\alpha \epsilon_{\alpha\beta}p^{(2)}_\beta) (k
\cdot \eta^{(1)})(k \cdot \eta^{(2)}) 
\end{equation} 
which would lead, in position space to a term like 
\begin{equation} 
(\partial_\mu
\partial_\nu h_{\alpha \beta})(\partial_\alpha X^\mu \partial_\beta
X^\mu) 
\end{equation} 
in the effective action. This would contribute to $l = 0$ and $l = 2$
partial wave absorptions. A detailed calculation shows that such terms
cancel between two groups of terms \cite{dasc} ! This result may
be also seen from the covariantized form of the expression for the
one loop effective action for four open string states given 
in \cite{Tseytlina}. 

The nonplanar diagram with two open string states involve only the
$U(1)$ piece. This cancels the planar diagram by the mechanism
described for diagrams with four open string states.

The lowest order contrbution in fact comes when the factors
$e^{ikX}$ are taken into account. This gives rise to an amplitude
of the form $g^2 \omega^6$. The corresponding cross-section is then
of the form $(gN)^4 \omega^{11}$ - {\em exactly what the black
hole ordered} !

\section{Conclusions}

The black hole-D brane correspondence requires that the open
string perturabtion theory of D-branes has a rather specific
structure with higher powers of couplings coming with specific
powers of the energy. We have found evidence for this structure
by looking at one loop processes explicitly. The results
presented above are, however, at best indicative of some deeper
structure of the theory.

A deficiency of our method is that strictly speaking we can
investigate only parallel p-branes. This is because of a lack of
detailed understanding of the microscopic model for bound states of
branes which describe five or four dimensional black holes in terms of
string diagrams. So far in these cases the effective string model has
been reasonably successful, though there are troublesome exceptions
like fixed scalars with $h_{55} \neq 0$ \cite{klebkrasb} (for other
discussions of the validilty of the effective string model see
\cite{klebmathur,traschen}) Furthermore higher angular momentum
emission seems to require an effective string tension $T_{eff}$ whose
origin is not understood \cite{mathurangular,gubser}. Recent work on
the moduli space of these models in the string theory \cite{wadia} as
well as in the M(atrix) theory \cite{dvv,martinec,halyo} may be useful in
this regard.

In fact, black holes may teach us important features of string
theory. Central to this is the issue of non-renormalization of
certain quantities which make the string theoretic models of
black hole work. An important example is the non-supersymmetric
but extremal holes \cite{Duff,dabholkar} where the agreeemnt
of the entropy may be explained by mass renormalization of these
states which are generically non-vanishing but vanish in the
limit of large mass \cite{dmr}.

We need a systematic understanding of the loop effects discussed
in this talk. After all it are these open string loop effects which
give rise to space-time structure and a good understanding is
certainly required to understand the physics of the horizon and
issues of information loss in terms of string physics.

\section{Acknowledgements} 

I would like to thank the organizers of
Strings 97, particularly R. Dijkgraff, E. Verlinde and H. Verlinde for
inviting me to this wonderful and stimulating conference and
A. Dabholkar, A. Jevicki, J. Maldacena, A. Sen, S. Shenker,
L. Susskind and S.R. Wadia for useful discussions and A. Tseytlin for
a correspondence. I especially thank S.D. Mathur and I. Klebanov for
enlightening discussions and sharing their insights.

\end{document}